\documentclass[conference,9pt]{IEEEtran}
\IEEEoverridecommandlockouts
\usepackage{cite}
\usepackage{amsmath,amssymb,amsfonts}
\usepackage{algorithmic}
\usepackage{graphicx}
\usepackage{textcomp}
\usepackage{xcolor}
\usepackage{hyperref}
\usepackage{mathrsfs}
\usepackage{booktabs}
\usepackage{array}

\def\BibTeX{{\rm B\kern-.05em{\sc i\kern-.025em b}\kern-.08em
    T\kern-.1667em\lower.7ex\hbox{E}\kern-.125emX}}

\newcommand{\papername}{DOMAC}
\newcommand{\delayimprovement}{6.5\%}
\newcommand{\areaimprovement}{25\%}

\author{
    \IEEEauthorblockN{
        Chenhao Xue\textsuperscript{1},
        Yi Ren\textsuperscript{1,2}, 
        Jinwei Zhou\textsuperscript{3}, 
        Kezhi Li\textsuperscript{4}, \\
        Chen Zhang\textsuperscript{5},
        Yibo Lin\textsuperscript{1,6,7},
        Lining Zhang\textsuperscript{8},
        Qiang Xu\textsuperscript{4,9},
        Guangyu Sun\textsuperscript{1,6,7,*}
        \thanks{\textsuperscript{*}Corresponding author: Guangyu Sun (\href{mailto:gsun@pku.edu.cn}{gsun@pku.edu.cn})}
    }
    \IEEEauthorblockA{
        \textsuperscript{1}\textit{School of Integrated Circuits},
        \textsuperscript{2}\textit{School of Software and Microelectronics, Peking University}, Beijing, China \\
        \textsuperscript{3}\textit{School of Integrated Circuits, Anhui Polytechnic University}, Wuhu, China \\
        \textsuperscript{4}\textit{Department of Computer Science and Engineering, The Chinese University of Hong Kong}, Sha Tin, Hong Kong S.A.R. \\
        \textsuperscript{5}\textit{Shanghai Jiao Tong University}, Shanghai, China \\
        \textsuperscript{6}\textit{Institute of Electronic Design Automation, Peking University}, Wuxi, China \\
        \textsuperscript{7}\textit{Beijing Advanced Innovation Center for Integrated Circuits}, Beijing, China \\
        \textsuperscript{8}\textit{School of Electronic and Computer Engineering, Peking University}, Shenzhen, China \\
        \textsuperscript{9}\textit{National Center of Technology Innovation for EDA}, Nanjing, China \\
        \{xch927027, yibolin, eelnzhang, gsun\}@pku.edu.cn, yiren20@stu.pku.edu.cn \\
        jwzhou0915@163.com, \{kzli24, qxu\}@cse.cuhk.edu.hk, chenzhang.sjtu@sjtu.edu.cn
    }
}

\begin{document}

\title{DOMAC: Differentiable Optimization for High-Speed Multipliers and Multiply-Accumulators}

\maketitle

\begin{abstract}

Multipliers and multiply-accumulators (MACs) are fundamental building blocks for compute-intensive applications such as artificial intelligence. With the diminishing returns of Moore's Law, optimizing multiplier performance now necessitates process-aware architectural innovations rather than relying solely on technology scaling. In this paper, we introduce \papername{}, a novel approach that employs differentiable optimization for designing multipliers and MACs at specific technology nodes. \papername{} establishes an analogy between optimizing multi-staged parallel compressor trees and training deep neural networks. Building on this insight, \papername{} reformulates the discrete optimization challenge into a continuous problem by incorporating differentiable timing and area objectives. This formulation enables us to utilize existing deep learning toolkit for highly efficient implementation of the differentiable solver. Experimental results demonstrate that \papername{} achieves significant enhancements in both performance and area efficiency compared to state-of-the-art baselines and commercial IPs in multiplier and MAC designs.

\end{abstract}


\section{introduction}
\label{sec:introduction}

In digital circuit design, multipliers and multiply-accumulators (MACs) constitute fundamental building blocks, playing a critical role in compute-intensive applications such as artificial intelligence and high-performance computing. As these modules play pivotal roles in determining the performance and efficiency of the entire hardware systems, optimizing for high-speed multipliers and MACs has become increasingly imperative.

Modern multipliers and MACs typically consist of three primary components: a partial product generator (PPG), a compression tree (CT), and a final carry-propagate adder (CPA). The CT realizes parallel compression of the partial product array into two rows, which are then processed by the CPA to produce the final result. Previous research has shown that CT accounts for more than 50\% of the delay and area in the entire multiplier, underscoring its critical importance in the overall design~\cite{feng2024gomarl}.

The optimization of high-speed CT has been a focal point of research for decades. 
Early advancements begin with classical structures such as Wallace tree~\cite{wallace1964suggestion} and Dadda tree~\cite{dadda1990some}, followed by numerous improvements that introduce compressor variants~\cite{nagamatsu199015,mehta1991high,song1991circuit} or customized architectures~\cite{bickerstaff1993reduced,fadavi1993m} optimized for specific technology nodes. However, these approaches heavily rely on domain expertise and require laborious redesign efforts when migrating to new technology nodes, which can significantly delay time-to-market.
Algorithmic solutions employ heuristic strategies and mathematical programming to automate the exploration of CT architecture variants. GOMIL adopts integer linear programming (ILP) to adjust compressor assignment for reduced area~\cite{xiao2021gomil}. The three-dimensional method proposes a heuristic approach to adjust compressor interconnection for balanced delay paths~\cite{oklobdzija1996method}. UFO-MAC leverages ILP to achieve global optimization of compressor interconnection, enhancing the overall speed~\cite{zuo2024ufo}. Nevertheless, these methods exhibit a significant limitation in oversimplifying the timing and area characteristics of compressors. Specifically, they assume fixed latency and area for compressors; whereas in practical logic synthesis, each compressor may be mapped to implementations with varying timing and area characteristics. Additionally, factors such as signal transition time and capacitive load can introduce timing variations, which are not fully accounted for. Consequently, the oversimplification may lead to suboptimal solutions for the targeted technology nodes.
More recently, reinforcement learning (RL) has been employed to search for physically-optimized multiplier and MAC designs~\cite{zuo2023rl,feng2024gomarl,lai2024scalable}. While delivering promising results, RL-based methods suffer from prohibitive computational complexity, as training RL agents requires extensive trial-and-error iterations and expensive invocations of synthesis tools during exploration.
In summary, the process-aware architectural optimization for high-performance multipliers and MACs urgently requires automated solutions that balance optimization efficiency and effectiveness.

To resolve the aforementioned limitations, we introduce an innovative solution that leverages \underline{d}ifferentiable \underline{o}ptimization to generate technology-specific high-speed multipliers and \underline{MAC}s, referred to as \papername{}. 
\papername{} evaluates the performance of CTs with detailed static timing analysis process (STA)~\cite{bhasker2009static}, and jointly manipulates compressor interconnection and technology mapping to improve post-synthesis timing and area. To efficiently address this large-scale combinatorial optimization problem, we establish an analogy between optimizing multi-staged CT and training deep neural networks (DNNs). Based on the analogy, \papername{} proposes differentiable timing and area objectives with respect to CT interconnection and implementation, which enables highly efficient optimization via gradient descent. 
The key contributions of this work are summarized as follows:

\begin{itemize}
    \item We propose \papername{}, a high-speed multiplier and MAC generation framework that jointly optimizes compressor interconnections and compressor physical implementations.   
    \item We analogize CT optimization with training DNNs and introduce differentiable timing and area objectives for efficient optimization of post-synthesis QoRs.
    \item We implement a highly efficient differentiable solver using established deep learning toolkit \texttt{PyTorch}~\cite{paszke2019pytorch}.
    \item Experiment results demonstrate that \papername{} optimized multipliers and MACs exceed all baseline designs, with up to \delayimprovement{} improvement in delay and \areaimprovement{} reduction in area over commercial IPs.
\end{itemize}

The rest of this paper is organized as follows: Section~\ref{sec:background} provides preliminaries of multiplier and MAC design. Section~\ref{sec:methodology} details \papername{} framework. Section~\ref{sec:experiment} presents the experimental results. Finally, Section~\ref{sec:conclusion} concludes this paper.

\section{preliminaries}
\label{sec:background}

\begin{figure}[!t]
  \centering
  \includegraphics[width=0.9\linewidth]{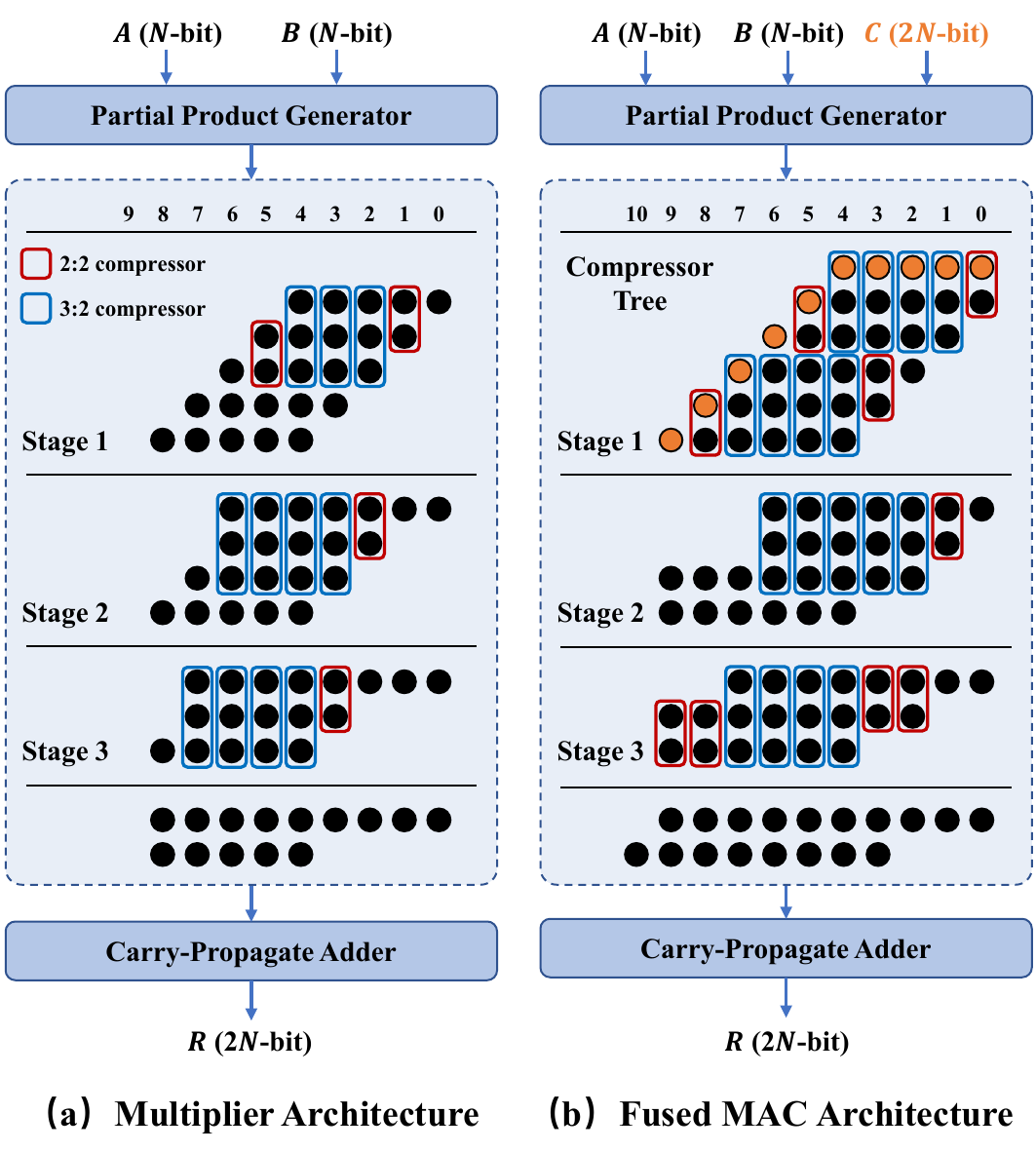}
  \caption{Architecture of (a) multipliers and (b) fused multipliy-accumulators}
  \label{fig:mult_arch}
\end{figure}

\subsection{Multiplier Architecture}

As shown in Fig.~\ref{fig:mult_arch}(a), a high-speed multiplier comprises three main components: a partial product generator (PPG), a compressor tree (CT), and a carry-propagate adder (CPA). 

\begin{itemize}

\item \textit{Partial Product Generator:} 
The PPG generates an array of partial products (PPs), where each PP is shifted into different columns to represent varied power-of-two weights.
A commonly employed \texttt{AND} gate-based PPG generates $N^2$ PPs for an $N$-bit multiplier.

\item \textit{Compressor Tree:}
The CT compresses PPs for multiple stages until a maximum of two PPs remain in each column.
CT typically consists of various types of compressors, with 3:2 and 2:2 compressors being predominantly used.

\item \textit{Carry-Propagate Adder:} 
The CPA aggregates the compressed PPs from the CT to produce the final product.
High-speed multipliers often utilize parallel prefix adders~\cite{sklansky1960conditional} for fast computation.

\end{itemize}

The fused multiply-accumulator (fused MAC) integrates the accumulation operation into the compressor tree to boost computation efficiency. As illustrated in Fig.~\ref{fig:mult_arch}(b), fused MAC exhibits a similar architecture as multiplier.

\subsection{Problem Formulation}

\papername{} focuses on exploring the design space of compressor trees (CTs), one of the most critical modules influencing the performance of multipliers and MACs. This section examines the comprehensive CT design space and formulates the design space exploration problem addressed by \papername{}.

\textbf{Compressor Tree Design Space} Factors that affect CT performance are as follows:

\begin{figure}[!t]
  \centering
  \includegraphics[width=\linewidth]{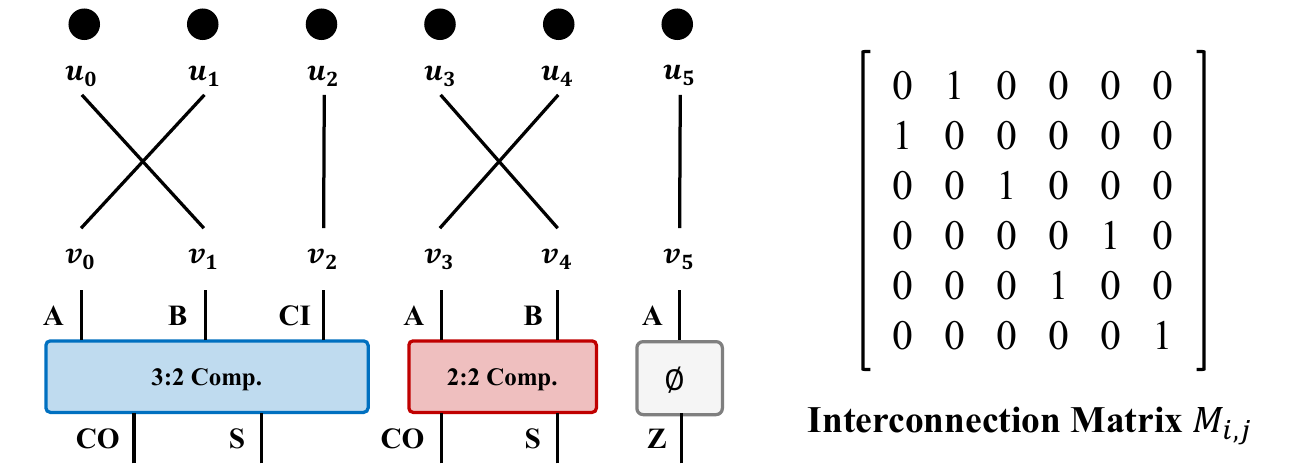}
  \caption{Example of adjusting compressor tree interconnection, where the partial products and compressor inputs can be assigned interchangeably.}
  \label{fig:ct_interconnection}
\end{figure}

\begin{figure}[!t]
  \centering
  \includegraphics[width=\linewidth]{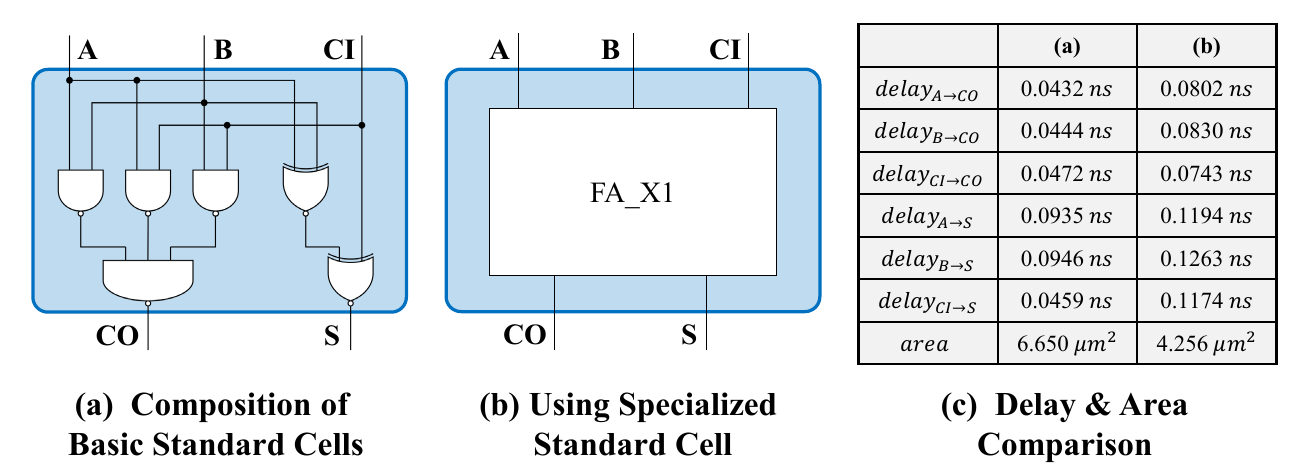}
  \caption{Example of two different implementations for 3:2 compressor using Nangate45 Open Cell Library~\cite{nangate2008freepdk45}. We estimate the delay using a nonlinear delay model (NLDM) with $\text{islew} = 0.02ns$ and $\text{oload} = 3fF$ for all input-output pairs.}
  \label{fig:compressor_implementation}
\end{figure}

\begin{itemize}

\item \textit{Compressor Type}: Different types of compressors have similar functions in reducing the number of partial products but differ in terms of the number of inputs or outputs. For a fair comparison with previous methods~\cite{wallace1964suggestion,xiao2021gomil,dadda1990some,lai2024scalable}, \papername{} utilizes commonly adopted 3:2 and 2:2 compressors to build CTs. However, \papername{} can be easily extended to support other types of compressors.

\item \textit{Compressor Quantity}: Each compression stage should comprise a proper amount of compressors. \papername{} utilizes a designated compressor assignment for subsequent optimization. Without loss of generality, \papername{} adopts the quantity derived from classical architectures such as Wallace tree~\cite{wallace1964suggestion} and Dadda tree~\cite{dadda1990some} .

\item \textit{Compressor Interconnection}: As shown in Fig.~\ref{fig:ct_interconnection}, the partial products and input ports of compressors located at column $i$ and stage $j$ must be mapped bijectively. Due to the associative property of addition, this mapping can be arbitrarily permuted without affecting the correctness of final results. We use binary matrix $M_{i,j}$ to represent the interconnection, where $M_{i,j}[u,v] = 1$ indicates that the $u$-th partial product is mapped to the $v$-th input port. Each row and column in matrix $M_{i,j}$ sums to 1, making $M_{i,j}$ a doubly stochastic matrix.

\item \textit{Compressor Implementation}: As shown in Fig.~\ref{fig:compressor_implementation}, a compressor may have different implementations with distinct timing and area characteristics. For each compressor $c$, we use one-hot vector $p_c$ to denote its physical implementation, in which $p_c[k] = 1$ indicates that $c$ adopts the $k$-th implementation from the available set $\mathcal{P}_c$.

\end{itemize}

\textbf{Problem} (\textit{Compressor Tree Design Space Exploration}) Given an initial compressor set $\textbf{C}$ where each compressor $c$ has been assigned to a specific column and stage, the objective is to minimize the absolute values of worst negative slack (WNS), total negative slack (TNS), and total area by simultaneously adjusting compressor interconnection $\textbf{M}$ and compressor physical implementation $\textbf{p}$. The formal definitions are presented as follows:

\begin{equation}
\begin{aligned}
    \min_{\textbf{M}, \textbf{p}} ~~~~ & \quad t_1 \text{WNS}(\textbf{M}, \textbf{p}) + t_2 \text{TNS}(\textbf{M}, \textbf{p}) + \alpha \text{Area}(\textbf{p}) \\
    \text{subject to} & \quad M_{i,j} ~ \text{is doubly stochastic}, ~ \forall M_{i,j} \in \textbf{M}; \\
                      & \quad \sum_k p_c[k] = 1, ~ \forall c \in \textbf{C}; \\
\end{aligned}
\label{eq:original_problem}
\end{equation}

The combinatorial optimization problem in Equation~\eqref{eq:original_problem} presents challenges for conventional discrete solvers. On one hand, the static timing analysis process is highly complex~\cite{bhasker2009static}, making it difficult to incorporate into standard solvers without substantial simplification. On the other hand, the problem involves numerous variables and intricate constraints, which render discrete optimization algorithms inefficient.

\section{Methodology}
\label{sec:methodology}

In this section, we propose the differentiable framework for optimizing technology-specific multipliers and MACs. Section~\ref{sec:inspiration} presents an inspiration from deep neural network. Section~\ref{sec:framework} presents an overview of the framework. Sections~\ref{sec:area_objective}, \ref{sec:timing_objective}, and~\ref{sec:constraint_objective} give the mathematical details of the area, timing, and constraint objectives. Section~\ref{sec:parameter} details the selection of hyperparameters.

\subsection{Inspiration from Deep Neural Networks}
\label{sec:inspiration}


Recent advancements in the theory and practice of machine learning have brought new inspiration for electronic design automation (EDA). Drawing parallels between EDA challenges and machine learning problems, researchers have developed novel algorithms to address analytical placement~\cite{lin2019dreamplace,guo2022differentiable} and gate sizing~\cite{du2024fusion}.
In \papername{}, we expand this approach to the efficient synthesis of technology-specific high-performance multipliers and MAC units.

As illustrated in TABLE~\ref{tab:analogy}, delay propagation in multi-staged compressor trees exhibits a striking analogy to the forward propagation process of deep neural networks (DNNs).
\begin{enumerate}
    \item The net delay propagation from a set of compressor output ports $\mathbf{u}$ to the corresponding set of compressor input ports $\mathbf{v}$ is determined by interconnection $M_{i,j}$. This process can be formulated as $\text{AT}(\mathbf{v}) = M_{i,j}^T \cdot \text{AT}(\mathbf{u})$, which resembles the linear projection $WX$ in DNNs.
    \item The cell delay propagation involves adding the internal cell delay to the arrival time at the compressor input ports, which is analogous to adding biases $b$ in DNNs.
    \item The timing objective WNS and TNS is computed by measuring the difference between final arrival time and required arrival time, which is similar to evaluating the consistency between DNN outputs and data labels through loss function $\mathcal{L}$.
\end{enumerate}



The analogy in TABLE~\ref{tab:analogy} inspires us to optimize CT design using a similar approach to training DNNs. In practice, DNNs are efficiently trained with backpropagation (BP), where weights $W$ and biases $b$ are updated using the gradient of loss function $\mathcal{L}$ w.r.t. these parameters. However, BP cannot be directly adopted due to fundamental distinction between the two problems: CT design involves discrete interconnections and physical implementations, whereas DNN parameters are continuous in nature. 

To address this challenge, we propose to relax the discrete CT optimization problem into a continuous formulation. In a nutshell, we relax the discrete selections of compressor interconnection $\mathbf{M}$ and compressor implementation $\mathbf{p}$ to continuous variables, making them essentially represent probabilistic distributions. Therefore, we can estimate the expectation of intermediate values in STA, such as signal arrival time and signal transition time (slew), and ultimately derive differentiable timing objectives. We use gradient descent to update $\mathbf{M}$ and $\mathbf{p}$ in continuous domain and then adopt legalization process to map them back to discrete domain. Consequently, a large portion of the CT optimization problem can be efficiently addressed following the same process as DNN training. Furthermore, we incorporate area objectives to balance the performance and area footprints, as well as constraint objectives to ensure the accuracy of timing and area estimations during continuous optimization. The details will be discussed in the following subsections.


\begin{table}[!t]
\centering
\caption{Analogy between optimizing compressor tree designs and training deep neural networks.}
\begin{tabular}{p{11em}<{\centering}|p{11em}<{\centering}}
    \toprule
    \textbf{DNN Training}    & \textbf{CT Optimization}    \\ \midrule
    Activations $X$ & Arrival time $AT$   \\
    Weights $W$     & Interconnection $M$ \\
    Biases $b$      & Cell delay $d$      \\
    Loss function $\mathcal{L}$  & Objective WNS, TNS \\ \bottomrule
\end{tabular}
\label{tab:analogy}
\end{table}

\subsection{Framework Overview}
\label{sec:framework}

The \papername{} framework consists of the following steps: 
\begin{enumerate}
    \item \textit{Differentiable Optimization:} Solve the continuous formulation of the CT optimization problem, which is essentially analogous to training DNNs. Leveraging the established deep learning toolkit \texttt{PyTorch}~\cite{paszke2019pytorch}, we efficiently implement our differentiable solver. 
    \item \textit{Legalization:} Map the continuous solution to a discrete solution. For each compressor interconnection $M_{i,j} \in \mathbf{M}$, we employ the Hungarian algorithm to identify the bipartite matching with the maximum probability summation. For each compressor implementation $p_c \in \mathbf{p}$, we apply the \texttt{argmax} operation to select the implementation with the highest probability.
    \item \textit{Netlist Generation:} Generate the synthesized gate-level CT netlists, which can be seamlessly integrated into multiplier and MAC designs.
\end{enumerate}

\subsection{Differentiable Area Objective}
\label{sec:area_objective}

We begin with the total area, which is the simplest objective. For each compressor $c$, the area of all available implementations is represented by a vector $a_c$. The area expectation can be derived from the probabilistic distribution $p_c$:
\begin{equation}
    \text{Area}_c(p_c) = \sum_k p_c[k] \cdot a_c[k]
\end{equation}
The total area of the compressor tree is the sum of the expected area across all compressors:
\begin{equation}
    \text{Area}(\mathbf{p}) = \sum_c \text{Area}_c(p_c)
\end{equation}
\subsection{Differentiable Timing Objective}
\label{sec:timing_objective}

Compared with area objectives, timing objectives are quite complex. The compressor implementation $\mathbf{p}$ affects the cell delay characteristics, whereas the interconnection $\mathbf{M}$ affects the delay propagation. Moreover, $\mathbf{p}$ and $\mathbf{M}$ jointly affect the capacitive loads, which further complicates the process of static timing analysis (STA). Next, we detail each stage of the differentiable STA in \papername{}:

\subsubsection{Pin Load Estimation}
\label{sec:pin_load_estimation}

Non-linear delay model (NLDM) is extensively adopted to characterize cell timing statistics in STA~\cite{bhasker2009static}. In NLDM, cell delay and slew are determined through look-up tables (LUTs). Given input slew and capacitive load of the cell, the delay and output slew are computed using linear interpolation or extrapolation based on a corresponding $2 \times 2$ submatrix within the LUT. Therefore, our first step is to derive cell capacitive load before timing propagation. Since both $\mathbf{M}$ and $\mathbf{p}$ represent probabilistic distributions, the exact values of capacitive load are inaccessible. Our best effort is to compute the expected capacitive load:
\begin{subequations}
\begin{align}
    \text{Cap}(v) &= \sum_k p_c[k] \cdot \text{Cap}(v; \mathcal{P}_c[k]) \\
    \text{Load}(u) &= \sum_{\text{child} ~ v} M_{i,j}[u,v] \cdot \text{Cap}(v)
\end{align}
\label{eq:expected_load}
\end{subequations}
where $\text{Cap(v)}$ denotes the expected capacitance at the input pin $v$ of the next-level compressor, and $\text{Load}(u)$ denotes the expected capacitive load at the output pin $u$ of previous-level compressor.

\subsubsection{Cell Delay Propagation}
\label{sec:cell_delay_propagation}

Next, we utilize NLDM to compute the delay, slew, and arrival times: 
\begin{subequations}
\begin{align}
    \text{Slew}_u(v)  &= \sum_k p_c[k] \cdot \text{LUT}^{\text{slew}}_{u \rightarrow v}(\text{Slew}(u), \text{Load}(v); \mathcal{P}_c[k]) \label{eq:cell_islew} \\
    \text{Delay}_u(v) &= \sum_k p_c[k] \cdot \text{LUT}^{\text{delay}}_{u \rightarrow v}(\text{Slew}(u), \text{Load}(v); \mathcal{P}_c[k]) \label{eq:cell_delay} \\
    \text{AT}(v)      &= \text{LSE}_\gamma \{ \text{AT}(u) + \text{Delay}_u(v) ~|~ \forall ~ \text{input} ~ u \} \label{eq:cell_at} \\
    \text{Slew}(v)    &= \text{LSE}_\gamma \{ \text{Slew}_u(v) ~|~ \forall ~ \text{input} ~ u \} \label{eq:cell_oslew}
\end{align}
\label{eq:cell_nldm}
\end{subequations}
In Equation~\eqref{eq:cell_islew} and Equation~\eqref{eq:cell_delay} , $u$ and $v$ are the input pin and output pin of compressor $c$, respectively. The slew and delay from $u$ to $v$ are estimated using the LUTs extracted from \texttt{.lib} files in the process design kit (PDK). In practice, the slew and delay for each output pin may correspond to multiple LUTs, each associated with a specific input assignment and signal edge (\texttt{rise} or \texttt{fall}). In \papername{}, we extract LUTs corresponding to the worst-case scenarios, where each entry in the LUT is selected from the maximum value across all possible conditions. Similar to Equation~\eqref{eq:expected_load}, we estimate the expectation of delay and slew under probabilistic implementation $p_c$. 

In Equation~\eqref{eq:cell_at} and Equation~\eqref{eq:cell_oslew}, we compute the arrival time and slew of pin $v$ by gathering the maximum value from all the related input $u$. We replace the $\texttt{max}$ operation with a smoothed approximation, the Log-Sum-Exp (LSE), as defined in Equation~\eqref{eq:lse}. This approach has been adopted by previous works to alleviate oscillation and instability in the differentiable optimization process~\cite{lin2019dreamplace,guo2022differentiable}. The hyperparameter $\gamma$ controls the trade-off between the smoothness and approximation accuracy.
\begin{equation}
    \text{LSE}_\gamma(x_1, x_2, \cdots, x_n) = \gamma \log(\sum_{i=1}^{n} \exp \dfrac{x_i}{\gamma})
\label{eq:lse}
\end{equation}

\subsubsection{Net Delay Propagation}

Next, we consider net delay propagation. In compressor trees where all nets exhibit limited fanouts and lengths, the slew variation and delay introduced by the nets themselves are negligible. Therefore, we mainly consider the impact of probabilistic interconnection $\mathbf{M}$ on net delay propagation. As discussed in Section~\ref{sec:pin_load_estimation}, we derive the expected slew and arrival time of next-level compressor input pin $v$:
\begin{subequations}
\begin{align}
    \text{Slew}(v) &= \sum_{\text{parent} ~ u} M_{i,j}[u,v] \cdot \text{Slew}(u) \\
    \text{AT}(v)   &= \sum_{\text{parent} ~ u} M_{i,j}[u,v] \cdot \text{AT}(u)
\end{align}
\end{subequations}
where the slew and arrival time of previous-level compressor output pin $u$ have been computed as detailed in Section~\ref{sec:cell_delay_propagation}.

\subsubsection{Timing Slack Estimation}

Finally, we derive the timing slacks as optimization objectives. For complex combinatorial logic such as compressor tree, our primary focus is to avoid setup timing violations, which occur when the setup slack becomes negative. Specifically, we aim to minimize the absolute value of worst negative slack (WNS) and total negative slack (TNS) defined as follows:
\begin{subequations}
\begin{align}
    \text{Slack}(u) &= \text{RAT}(u) - \text{AT}(u) \\
    \text{WNS}(\mathbf{M}, \mathbf{p}) &= \text{LSE}_\gamma\{\min(0, -\text{Slack}(u)) | ~\text{output} ~ u\} \\
    \text{TNS}(\mathbf{M}, \mathbf{p}) &= \sum_{\text{output} ~ u} \min(0, -\text{Slack}(u))
\end{align}
\end{subequations}
where $u$ is the primary output pin of the compressor tree. $\text{RAT}(u)$ is the required arrival time for pin $u$, assigned by the designer. We replace the \texttt{max} operation in $\text{WNS}$ with LSE for better smoothness, following the same discussion in Section~\ref{sec:cell_delay_propagation}.

\subsection{Optimization Constraints}
\label{sec:constraint_objective}

Given the relaxed continuous version of the problem defined in Equation~\eqref{eq:original_problem}, we aim to ensure the differentiable optimization yields valid discrete solutions. This is achieved through variable substitution and the introduction of constraint loss functions.

\subsubsection{Variable Substitution}

For a valid probabilistic distribution of compressor implementation $p_c$, the summation of all elements must equal 1.
To satisfy this constraint, we introduce auxiliary variables $\tilde{p}_c \in \mathbb{R}^{|\mathcal{P}_c|}$ and compute $p_c$ using the \texttt{softmax} function:
\begin{equation}
    p_c = \texttt{softmax}(\tilde{p}_c) = [\dfrac{e^{\tilde{p}_c[0]}}{\sum_k e^{\tilde{p}_c[k]}}, \cdots, \dfrac{e^{\tilde{p}_c[|\mathcal{P}| - 1]}}{\sum_k e^{\tilde{p}_c[k]}}]
\end{equation}
For the doubly stochastic matrix $M_{i,j}$ representing compressor interconnection, we adopt a similar approach to ensure the sum of each row equals 1. Specifically, we introduce auxiliary variables $\tilde{M}_{i,j} \in \mathbb{R}^{l \times l}$, where $l$ is the number of partial products at column $i$ and stage $j$. The matrix $M_{i,j}$ is computed as:
\begin{equation}
    M_{i,j}[u,\cdot] = \texttt{softmax}(\tilde{M}_{i,j}[u,\cdot])
\end{equation}
Variable substitution guarantees that the probabilistic constraints are consistently upheld while permitting the auxiliary variables to be freely adjusted.

\subsubsection{Bijective Mapping Loss}

The doubly stochastic property of $M_{i,j}$ also demands the sum of each column equals 1, which is fulfilled by introducing the bijective mapping loss in Equation~\eqref{eq:bm_loss}. The quadratic form is chosen for better smoothness.
\begin{equation}
    \mathcal{L}_{\text{BM}}(\mathbf{M}) = \sum_{i,j,u} (\sum_{v} M_{i,j}[u,v] - 1)^2
\label{eq:bm_loss}
\end{equation}

\subsubsection{Discretization Loss}

To encourage continuous variables to converge to bimodal states (0 or 1), we introduce the discretization loss in Equation~\eqref{eq:d_loss}. This function is minimized at 0 and 1 and exhibits smooth gradients near these minima:
\begin{subequations}
\begin{align}
    \mathcal{L}_{\text{D}}(x) &= x^2 \cdot (1-x)^2 \\
    \mathcal{L}_{\text{D}}(\mathbf{M},\mathbf{p}) &= \sum_{i,j,u,v} \mathcal{L}_{\text{D}}(M_{i,j}[u,v]) + \sum_{c,k} \mathcal{L}_{\text{D}}(p_{c}[k])
\end{align}
\label{eq:d_loss}
\end{subequations}

\subsection{Hyperparameter Selection}
\label{sec:parameter}

In the previous sections, we have reformulated the discrete CT optimization problem into a differentiable optimization problem. The complete objective function is summarized as follows:
\begin{equation}
\begin{aligned}
    \min_{\tilde{\mathbf{M}}, \tilde{\mathbf{p}}} ~~~~~ & t_1 \text{WNS}(\mathbf{M}, \mathbf{p}) + t_2 \text{TNS}(\mathbf{M}, \mathbf{p}) + \alpha \text{Area}(\mathbf{p}) \\
    + & \lambda_1  \mathcal{L}_{D}(\mathbf{M}, \mathbf{p}) + \lambda_2 \mathcal{L}_{BM}(\mathbf{M})
\end{aligned}
\end{equation}
where $t_1, t_2, \alpha, \lambda_1, \lambda_2$ are weights for different objectives. Without loss of generality, we assume the timing slacks are measured in nanoseconds $ns$ and area are measured in square micrometers $\mu m^2$. 
The optimization process is conducted over 300 iterations, with incremental adjustments to the hyperparameters starting from the 100th iteration. This approach ensures a balanced trade-off among the objectives at different stages of the optimization. Specifically, 
$\alpha$ is set between $1$ and $5$ to trade-off timing and area, and is increased by 0.3\% every iteration to counterbalance timing objectives.
$t_1$ and $t_2$ are set to 1 and 0.01, respectively, and are increased by 0.5\% every iteration to prioritize timing optimization in the later stages. We set required arrival time $\text{RAT}(u) = 0$ for all output ports $u$.
$\lambda_1$ and $\lambda_2$ are set to 0.1 and 0.5, respectively, and are increased by 1\% per iteration to ensure that design constraints are adequately respected without excessively disrupting the optimization process.
The smoothing factor $\gamma$ for $\text{LSE}$ operations is set to 0.01, which provides sufficient approximation accuracy and smoothness for differentiable optimization.

\section{experiment results}
\label{sec:experiment}
\subsection{Experiment Setup}

All experiments are conducted on a Linux-based platform equipped with 2.8GHz 96-core Intel Xeon Gold 6342 CPUs and 1.5TB memory. Unless specified otherwise, all multipliers and MACs adopt \texttt{AND}-based PPG and default CPA instantiated with \texttt{s=a+b} style RTL. We compare the multipliers and MACs generated by \papername{} with several baseline approaches, including: Wallace trees~\cite{wallace1964suggestion}, Dadda trees~\cite{dadda1990some}, and commercial IPs from Synopsys DesignWare Library~\cite{SynopsysDW}. The commercial IPs are instantiated with \texttt{r=a*b} and \texttt{r=a*b+c} style RTL. Additionally, the multipliers are compared with those produced by GOMIL~\cite{xiao2021gomil} (ILP-based) and ArithmeticTree~\cite{lai2024scalable} (RL-based) leveraging their open-sourced implementation.
All designs are synthesized by Synopsys Design Compiler (Version R-2020.09-SP2)~\cite{SynopsysDC} using Nangate 45nm Open Cell Library~\cite{nangate2008freepdk45} and the \texttt{compile\_ultra} command. To illustrate the trade-off between delay and area, we sweep the target delay constraints from 0$ns$ to 2$ns$.

\subsection{Results and Analysis}



\begin{figure}[!t]
  \centering
  \includegraphics[width=\linewidth]{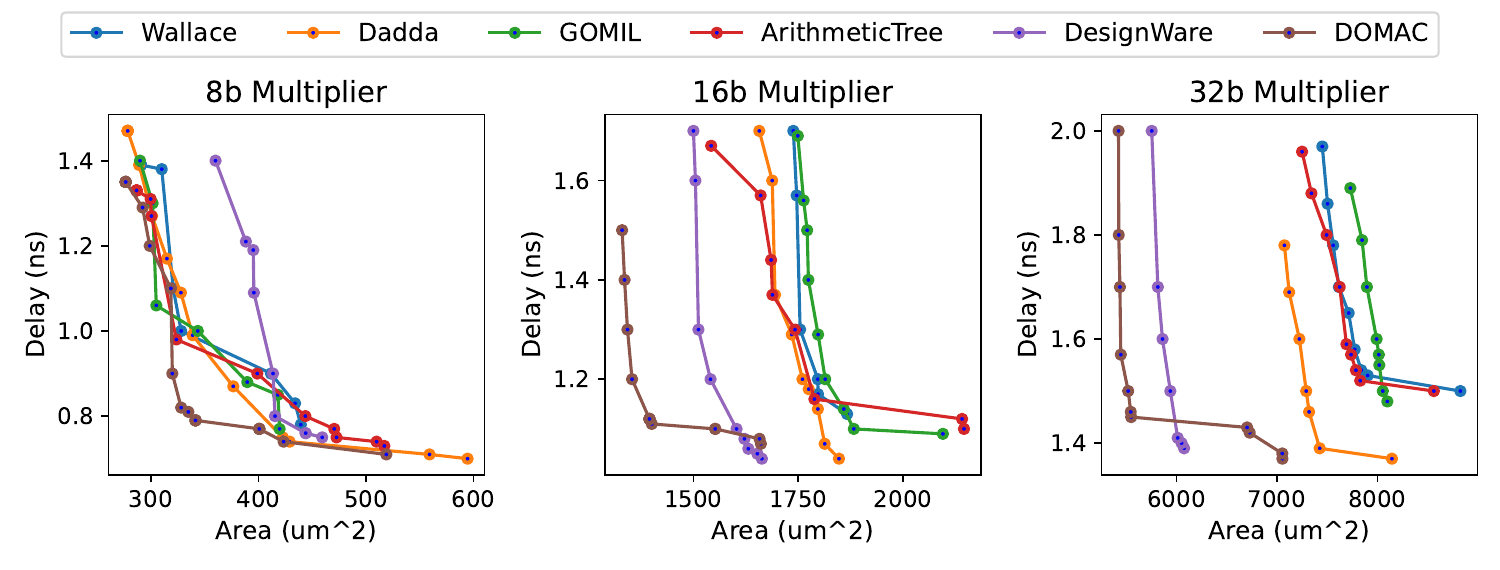}
  \caption{Pareto frontiers of the synthesized results on multipliers.}
  \label{fig:mult_pareto}
\end{figure}

\begin{figure}[!t]
  \centering
  \includegraphics[width=\linewidth]{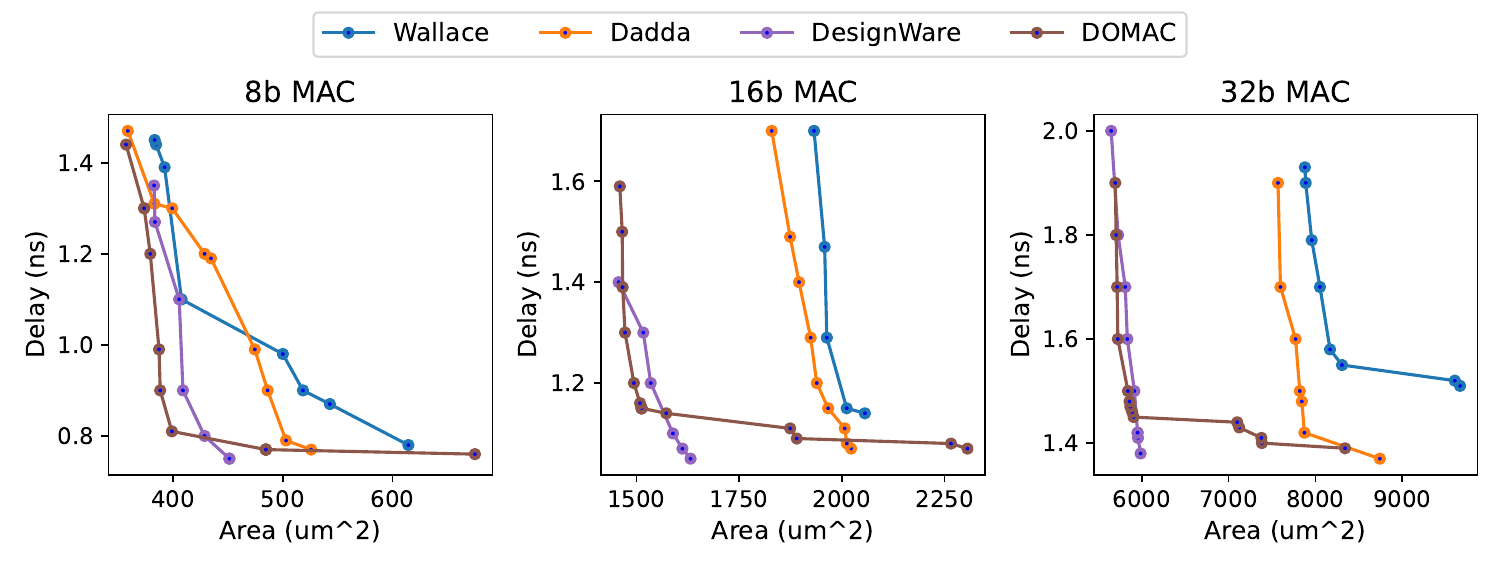}
  \caption{Pareto frontiers of the synthesized results on MACs.}
  \label{fig:mac_pareto}
\end{figure}

Fig.~\ref{fig:mult_pareto} illustrates the synthesized results of multipliers, displaying the Pareto frontiers with respect to delay and area. In most cases, \papername{} achieves superior performance over all baseline methods. Our framework identifies designs that Pareto-dominate those of the commercial IPs, delivering up to 6.5\% reduction in delay and 25\% reduction in area. GOMIL produces inferior results than \papername{}, as it does not account for physical implementation variations or interconnection orders. ArithmeticTree fails to effectively generate new Pareto-optimal designs, suggesting that the reinforcement learning agents struggle to explore the extensive design space under constrained computational budgets.
Fig.~\ref{fig:mac_pareto} presents the synthesis results of multiply-accumulators. Consistent with the trends observed in Fig.~\ref{fig:mult_pareto}, \papername{} demonstrates superior area efficiency to commercial IPs under relaxed timing constraints and delivers competitive performance when optimized for high-speed operation.

\begin{figure}[!t]
  \centering
  \includegraphics[width=0.9\linewidth]{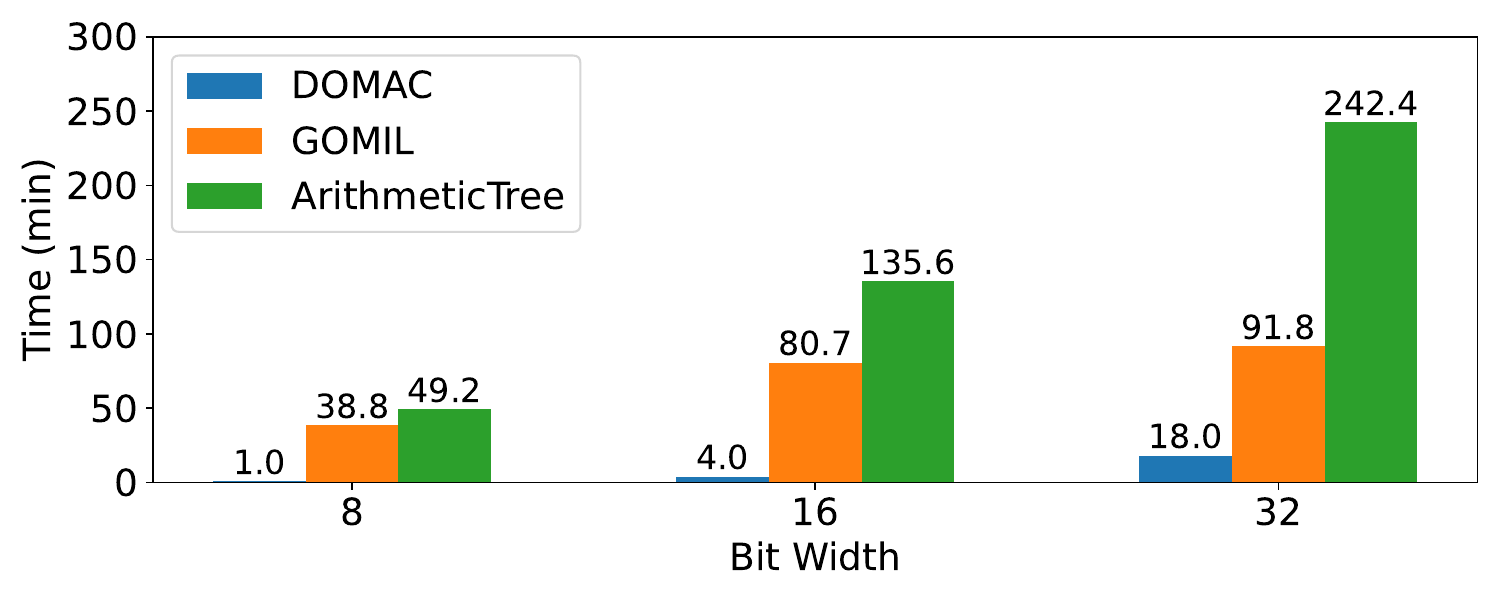}
  \caption{Runtime comparison between different methods.}
  \label{fig:runtime}
\end{figure}

We also evaluate the runtime of \papername{} for optimizing multiplier designs, which is shown in Fig.~\ref{fig:runtime}. For all bit width configurations, \papername{} requires less than 30 minutes, which is more efficient than the ILP-based approach GOMIL and RL-based method ArithmeticTree. The enhanced efficiency of \papername{} can be primarily attributed to two key factors. First, the differentiable formulation of the original CT optimization problem enables more efficient resolution compared to its discrete counterparts. Second, our differentiable solver takes advantage of the well-established automatic differentiation engine from the \texttt{PyTorch} toolkit.

\section{conclusion}
\label{sec:conclusion}

This paper introduces \papername{}, a differentiable optimization framework designed to synthesize high-speed multipliers and MACs. The core insight lies in establishing a parallel between compressor tree optimization and DNN training, thereby enabling efficient differentiable optimization for compressor trees. In future work, we plan to leverage GPU acceleration to further enhance optimization efficiency and extend the methodology to other critical components of multipliers, such as partial product generators and carry-propagate adders.

\section*{Acknowledgement}
This work is supported in part by Beijing Natural Science Foundation (Grant No. L243001), National Natural Science Foundation of China (Grant No. 62032001), and 111 Project (B18001).

{
\bibliographystyle{IEEEtran}
\bibliography{./references}
}

\end{document}